\input harvmac


\Title{\vbox{\baselineskip12pt\hbox{hep-th/9802116}
\hbox{LBNL-41416}
 \hbox{NSF-ITP-98-020}
\hbox{UCB-PTH-98/11}
}}
{\vbox{\centerline{Spectrum of Large $N$ Gauge Theory
from Supergravity} }}

\centerline{Gary T. Horowitz$^1$ and Hirosi Ooguri$^{2,3,4}$}
\bigskip\centerline{$^1$ Department of Physics, University
of California}
\centerline{Santa Barbara, CA 93106-9530}
\medskip
\centerline{$^2$ Department of Physics, University of California} 
\centerline{Berkeley, CA 94720-7300}
\medskip
\centerline{$^3$ Ernest Orlando Lawrence Berkeley National Laboratory}
\centerline{Mail Stop 50A-5101, Berkeley, CA 94720}
\medskip
\centerline{$^4$ Institute for Theoretical Physics, 
University of California}
\centerline{Santa Barbara, CA 93106}


\vskip .3in
\centerline{\bf Abstract}

\medskip
Recently, Maldacena proposed that the large $N$ 
limit of the ${\cal N}=4$ supersymmetric gauge theory 
in four dimensions with $U(N)$ gauge group
is dual to the type IIB superstring theory on 
${\rm AdS}_5 \times S^5$. 
We use this proposal to study the spectrum of the
large $N$ gauge theory on ${\bf R} \times S^3$ 
in a low energy regime.
We find that the spectrum is discrete and evenly spaced, 
and the number of states at each energy level 
is smaller than the one predicted by
the naive extrapolation of the Bekenstein-Hawking formula
to the low energy regime. We also show that 
the gauge theory describes a region of spacetime
behind the horizon as well as the region in front. 

\vfill\eject

\newsec{Introduction}

The number of 
states in a $d$-dimensional conformal 
field theory (CFT) grows as a 
function of the energy $E$ as
\eqn\scale{  S \sim E^{d-1 \over d} V^{1 \over d}, }
where $V$ is the volume of 
the $(d-1)$-dimensional space. 
For example, for $d=2$, the precise formula is 
$$ S_{d=2} = \sqrt{{2\pi \over 3} c E V}, $$
where $c$ is the central charge of the 
Virasoro algebra \ref\cardy{J. A. Cardy,
Nucl. Phys. B270 (1986) 186.}.  For $d=4$, 
the entropy of the ${\cal N}=4$  $U(N)$ gauge theory 
in the strong coupling regime is believed to be
\ref\gkp{S. S. Gubser, I. R. Klebanov and 
A. W. Peet, Phys. Rev. D54 (1996) 3915; A. Strominger,
unpublished.}
\eqn\threebrane{ S_{d=4} = (2/3)^{3 \over 4}
\sqrt{2 \pi N}
 E^{3 \over 4} V^{1 \over 4}}
because of the 
Bekenstein-Hawking entropy 
formula for the near extremal $3$-brane solution of type 
IIB theory \ref\hs{G. T. Horowitz and A. 
Strominger, Nucl. Phys. B360 (1991) 197.}.
In general \scale\ can be derived by 
assuming that the entropy is an extensive 
quantity and that it is invariant under the dilatation. 

Recently Maldacena made an interesting proposal in \ref\malda{J. 
Maldacena, hep-th/9711200, to appear 
in Adv. Theor. Math. Phys.} that compactifications 
of M/string theory on a sphere to  
anti-de \thinspace Sitter space (AdS) 
are dual to various conformal field theories.
In particular the large $N$ limit of the ${\cal N}=4$ $U(N)$ gauge 
theory in four dimensions is claimed to be described 
by the type IIB string theory 
on ${\rm AdS}_5 \times S^5$. This implies that spectra
of the two theories and hence the entropies are the same. 

The purpose of this paper is to use this proposal 
to learn about the spectrum of the large $N$ gauge theory
on ${\bf R} \times S^3$. We will establish a correspondence
between states in this gauge theory and states
in string theory on ${\rm AdS}_5 \times S^5$. 
In the supergravity approximation to
string theory, the energy levels 
are quantized in the units of the AdS radius
$R = (4 \pi gN)^{1 \over 4} l_s$ where $g$ is the string coupling
constant and $l_s$ is the string length,
$$ E = {n \over  (4 \pi gN)^{1 \over 4} l_s}, ~~~
(n=0,1,2, \cdots) .$$
The supergravity approximation is valid when $E$ is less than
both the string scale and the Planck scale, 
$$ E \ll l_s^{-1}, l_p^{-1} .$$
Since $l_p \sim g^{1 \over 4} l_s$ in ten dimensions, we can
trust the supergravity computation for 
\eqn\validity{ n \ll (gN)^{1 \over 4}. }
In this regime, we find that the entropy of the type IIB
string theory scales as a function of $E = n/R$ as
\eqn\sugra{ S_{AdS_5 \times S^5} \sim 
n^{9 \over 10} , }
for $n \gg 1$.

We will show that each supergravity state with
energy $E=n/R$ corresponds to a state with
energy $E=n$ in
the gauge theory on ${\bf R} \times S^3$ 
with a unit radius sphere. Thus Maldacena's
proposal leads to the prediction that the entropy of
the large $N$ gauge theory on ${\bf R} \times S^3$ for 
$1 \ll n \ll (gN)^{1 \over 4}$
is 
\eqn\gauge{ S_{gauge} \sim n^{9 \over 10}.}
In particular, the large $N$
spectrum is independent of $N$.

On the other hand, the
Bekenstein-Hawking formula \threebrane\ at this energy
reads
\eqn\bh{ S_{d=4} \sim \sqrt{N} n^{3 \over 4}. }
Since this formula was originally derived for the $3$-branes
wrapped on $T^3$, it is a prediction for the density of states for
the gauge theory on $T^3$. However for sufficiently large 
$n$, finite size effects are irrelevant and 
we expect that the density of states is independent of
the topology of the $3$-manifold. In such a case,
 we can use the formula \bh\
for the gauge theory on $S^3$ also. 
Comparing this with \gauge, the power of $n$
is different and there is no factor of $\sqrt{N}$  
in \gauge . 
Therefore in the low energy regime \validity , 
$S_{gauge}$
is smaller than the Bekenstein-Hawking entropy \threebrane. 
This does not mean that the proposal of \malda\ is wrong since
the finite size effects may become relevant for
$n \ll N^2$, which includes the low energy regime
we study in this paper\foot{
The finite size effects are negligible when the
wave length corresponding to temperature $T$ is much
shorter than the size of the $3$-manifold. For the $3$-branes,
$T \sim (n/N^2)^{1\over 4} V^{-{1\over 3}}$, so the finite
size effects are negligible for $n \gg N^2$.}. 

We also point out that Maldacena's conjecture implies that the four
dimensional gauge theory describes a region of spacetime {\it behind}
the horizon as well as the region in front.

\newsec{Conformal Symmetry and Spectrum}

${\rm AdS}_{d+1}$ has an isometry group of
$SO(2,d)$. In \malda , 
this group was identified
with the conformal symmetry of a gauge
theory on
${\bf R}^{1,d-1}$ for $d=3,4,6$. The conformal symmetry 
is generated 
by the momentum $P_\mu$,
the Lorentz generators $L_{\mu\nu}$, the dilatation
$D$ and 
the special conformal generators
$K_\mu$ ($\mu, \nu = 0, 1, \cdots ,d-1$). To simplify 
the following analysis, the generators are normalized
so that the AdS radius $R$ does not show up in the
structure constants.

Minkowski space ${\bf R}^{1,d-1}$ can be conformally 
embedded in the Einstein static universe, which has the topology 
of ${\bf R} \times S^{d-1}$, as follows. Introducing the null coordinates
$u=t - r$ and $v=t+r$, the
Minkowski metric is
$$ ds^2 = -du dv + {1 \over 4} (v-u)^2 d\Omega_{d-2} .$$
Multiplying this by the conformal factor, $4
(1+u^2)^{-1}(1+v^2)^{-1}$, and introducing new coordinates
$$ u = {\rm tan}\left({\tilde{t}-\chi \over 2}\right), 
~~v= {\rm tan}\left({\tilde{t}+\chi\over 2} \right) $$
the rescaled metric becomes 
$$  d\tilde{s}^2 = -d\tilde{t} + d\chi^2 + \sin^2\chi 
\  d\Omega_{d-2}, $$
which can be recognized as that of the Einstein universe
with unit radius. 
Although the original Minkowski space is mapped to 
a subset of this space, it was shown 
by L\"uscher and Mack \ref\lm{M. L\"uscher and G. Mack, 
Commun. Math. Phys. 41 
(1975) 203.} that correlation functions of 
CFT on ${\bf R}^{1,d-1}$ can be analytically 
continued to the full Einstein universe. Moreover, since
$$ {\partial \over \partial \tilde{t}} =
   {1 \over 2}(1 + u^2) {\partial \over \partial u} 
  + {1\over 2}(1 + v^2) {\partial \over \partial v},$$
the generator $H$
of the global time translation on ${\bf R} \times S^{d-1}$ is given by 
\eqn\hamilton{ H = {1\over 2}(P_0 + K_0).}
Students of two-dimensional CFT would recognize that
this is a higher dimensional generalization of the fact that
the Virasoro generator
$L_0$ is the translation generator on ${\bf R} \times S^1$ 
while the momentum on ${\bf R}^{1,1}$ is $L_{-1}$. 

${\rm AdS}_{d+1}$ is a globally static space and 
$H={1 \over 2}(P_0 + K_0)$ is also its global time translation generator.
Therefore the conjecture of \malda\ implies that 
each state of the CFT on ${\bf R} \times S^{d-1}$ is identified
with a state in the M/string theory on ${\rm AdS}_{d+1}$
times a sphere.  

The Hilbert space of the semi-classical supergravity
is constructed from a free gas of local fluctuations of 
the fields. 
Since AdS$_{d+1}$ is not globally hyperbolic (infinity is
a timelike boundary), we need to impose appropriate
boundary conditions on the supergravity fields. 
The requirement that the local fluctuations should give 
unitary representations of $SO(2,d)$ severely limits
the choice of boundary conditions \ref\ais{S. J. Avis, C. J. Isham
and D. Storey, Phys. Rev. D18 (1978) 3565.},\ref\fr{P. Breitenlohner
and D. Z. Freedman, Ann. Phys. 144 (1982) 249.}.  This is a reasonable
requirement in the present case as we expect that the CFT discussed
in \malda\ to have a unitary spectrum. 

Let us examine the case of the IIB theory on
${\rm AdS}_5 \times S^5$. The compactification of the supergravity
on $S^5$ creates a tower of massive particles on ${\rm AdS}_5$ and
their spectrum is classified in
\ref\gm{M. G\"unaydin and N. Marcus,  
Class. Quant. Grav. 2 (1985) L11.} and
 \ref\krn{H.J. Kim, 
L.J. Romans
and
P. van Nieuwenhuizen, Phys. Rev. D32 (1985)
389.}.  
Curiously they found that all the eigenvalues of $H$
are integers, including excitations on ${\rm AdS}_5$.
Even though AdS is a non-compact space, the curvature
introduces an effective infrared cutoff, which makes
the spectrum discrete.   
%
%
Since our $H$ is normalized so that the AdS radius $R$
does not show up in the structure constants of $SO(2,4)$, 
this means that all the wave modes are periodic in time with
the period $2 \pi R$ and the supergravity theory is 
well-defined in the single cover of ${\rm AdS}_5$.
(To be precise, the fermions have half odd integral modes
and therefore they are anti-periodic on ${\rm AdS}_5$.) We will
discuss implications of this observation to the near horizon
geometry of the $3$-brane later. 

The number of the Kaluza-Klein modes with  
$({\rm mass})^2 \sim (l/R)^2$ is of order $l^4$
for large $l$. The fluctuations
of each Kaluza-Klein mode on ${\rm AdS}_5$
gives a representation 
of $SO(2,4)$ with the
highest weight $H  \sim l$. The action 
of $SO(2,4)$ creates states with energy
$H \sim l + s$ $(s = 0, 1, 2, \cdots )$ with the asymptotic
degeneracy of order $s^3$, except for
a special representation called the 
singleton for which the degeneracy
grows slower \ref\fronsdal{C. Fronsdal, Rev. Mod. Phys. 37 (1965) 221;
Phys. Rev. D10 (1974) 589; Phys. Rev. D12 (1975) 3819;
C. Fronsdal and R. B. Haugen, Phys. Rev. D12 (1975) 3810.}.
The singletons appear
for special values of Kaluza-Klein masses
and do not contribute to the leading 
behavior of the entropy. The number
$\rho(\epsilon)$ of single particle states 
with the energy $H = \epsilon$ is therefore
\eqn\single{ \rho(\epsilon) \sim \sum_{(l, s); l+s = \epsilon} l^4 s^3
  \sim \epsilon^8. }
Following the standard procedure, the single particle spectrum
$\rho(\epsilon)$ can be converted into the entropy of a free
gas of particles on ${\rm AdS}_5 \times S^5$ with total energy
$H=n$ to obtain
\eqn\sphere{ S \sim n^{9 \over 10}. }
Since we identify
the AdS Hamiltonian $H={1 \over 2}(P_0 + K_0)$ with
the CFT Hamiltonian on ${\bf R} \times S^3$ with a unit radius, 
\sphere\ also gives the entropy of the CFT on ${\bf R} \times S^3$
at energy $E = n$.

It is straightforward to repeat the analysis for other 
supergravity background such as ${\rm AdS}_7 \times S^4$
and ${\rm AdS}_4 \times S^7$. In both cases, the entropy
behaves as 
$$ S \sim n^{10 \over 11} , $$
just as a free gas in eleven dimensions. 

There is a possibility that the correspondence of the 
large $N$ gauge theory and the supergravity on ${\rm AdS}_5 
\times S^5$ requires a non-standard choice 
of boundary conditions. In the above analysis
we only assumed that fluctuations of the fields 
make unitary representations of $SO(2,4)$, 
and the above estimate would not be sensitive 
to the precise choice of the boundary conditions
having such a property. 

The result \sphere\
has an obvious interpretation. 
When the radius of ${\rm AdS}_5 \times S^5$
is large, the space looks almost like ${\bf R}^{1,9}$.
The entropy $S \sim n^{9 \over 10}$ simply reflects
the bulk degrees of freedom in ten dimensions. 

\newsec{Spacetime Beyond the Horizon}

  The role of spacetime beyond the horizon has been puzzling, in 
  the recent successful description of
black hole microstates in terms of states of weakly coupled 
D-branes \ref\peet{For a recent review, see A. Peet, hep-th/9712253.}.
Maldacena's conjecture sheds light on this issue. The supergravity solution
describing $N$ extremal three branes consists of a completely nonsingular
spacetime with an infinite number of asymptotically flat regions, each with
a horizon \ref\ght{G.W. Gibbons, G.T. Horowitz and P.K. Townsend,
 Class. Quant. Grav. 12 (1995) 297.}.
One can periodically identify this space so that there are only
two asymptotically flat regions (one on each side of the horizon),
but then one introduces closed timelike
curves. The region near the horizon is, of course, locally 
${\rm AdS}_5$.
If one identifies to have only two asymptotically flat regions, the near
horizon geometry is globally ${\rm AdS}_5$.

Now consider
the four dimensional large $N$ gauge theory. The conformal symmetries of this
theory are globally either $SO(2,4)$ or its covering group.  These symmetries
must be reflected in the near horizon AdS geometry. The region to one
side of one horizon is not invariant under this group. One must include
at least the region on both sides of the horizon.  Thus {\it
the gauge theory describes spacetime on both sides of the horizon}.
The question of whether
the infinite chain of horizons must be included depends on whether the 
conformal
weights of the theory are all integer. 
If so, the conformal group is just $SO(2,4)$ (not the covering group)
and the spacetime contains just one horizon. We have seen that
supergravity on ${\rm AdS}_5 \times S^5$  
has integer energies with respect to the
global time, so it is well defined on the single cover of AdS$_5$.
 
The fact that the entire tower of
Kaluza-Klein states of supergravity have integer energy levels
can be interpreted in terms of the AdS supersymmetry algebra
\ref\fn{D. Z. Freedman and H. Nicolai, Nucl. Phys. B237 (1984) 342.}
\foot{We thank Paul Townsend for pointing this out to us.}. They are
BPS states, forming short supersymmetry multiplets, and their levels
are naturally integral. Assuming the correspondence of the AdS
supergravity and the CFT \malda, these Kaluza-Klein states should
correspond to chiral states in the gauge theory without anomalous
dimensions. By counting all
chiral states of the gauge theory, one should be able to check
the conjecture (1.5). On the other hand, there is no supersymmetry
reason to expect that massive string states have integral energy
levels. This suggests that the string theory should be defined
on the universal cover of AdS$_5$. 

After submitting this paper, we 
received a paper by Gubser, Klebanov and Polyakov
\ref\gkp{S.S. Gubser, I.R. Klebanov, and A.M. Polyakov, hep-th/9802109.}
where they suggested that 
the massive string states do not have integer energy levels and 
the anomalous dimensions of the 
corresponding operators
in the gauge theory are increased 
by the factor $(gN)^{1 \over 4}$. 
We also received a paper by Witten 
\ref\witten{E. Witten, hep-th/9802150.} where he outlined
the correspondence between some
chiral states in the gauge theory and the Kaluza-Klein 
supergravity states, extending earlier work 
\ref\klebanov{
I.R. Klebanov, Nucl. Phys. B496 (1997) 231; 
S.S. Gubser, I.R. Klebanov and A.A.Tseytlin,
Nucl. Phys. B499 (1997) 217.}. More recently
this correspondence has been established
for all supergravity states by
Ferrara, Fronsdal and Zaffaroni
\ref\ffz{S. Ferrara, C. Fronsdal
and A. Zaffaroni, hep-th/9802203.},
confirming the prediction of this paper. 

\bigbreak\bigskip\bigskip

\centerline{\bf Acknowledgments}
\nobreak
We would like to thank Juan Maldacena for discussions. 
The research of G.H. is supported in part by NSF grant
PHY95-07065. 
The research of H.O. is supported in part by NSF grant PHY95-14797 and
by DOE grant DE-AC03-76SF00098, and also by NSF grant PHY94-07194 
through the Institute for Theoretical Physics.

\listrefs

\end